\documentclass[aps,prd,longbibliography,twocolumn]{revtex4-2}
\usepackage[utf8]{inputenc}
\usepackage{graphicx}
\usepackage{bm}
\usepackage {amsmath}
\usepackage{color}

\begin{document}

\preprint{KUNS-2879, YITP-21-62, OCU-PHYS-542, AP-GR-170}

\title{Oscillations in the EMRI gravitational wave phase correction as a probe of reflective boundary of the central black hole}
\author{Norichika Sago$^{a,b}$}
\author{Takahiro Tanaka$^{a,c}$}
\affiliation{$^a$Graduate School of Science, Kyoto University, Kyoto 606-8502, Japan}
\affiliation{$^b$Advanced Mathematical Institute, Osaka City University, Osaka 558-8585, Japan}
\affiliation{$^c$Center for Gravitational Physics, Yukawa Institute for Theoretical Physics, Kyoto University, Kyoto 606-8502, Japan}
\date{\today}

\begin{abstract}
We discuss the energy loss  
due to gravitational radiation of binaries composed of exotic objects 
whose horizon boundary conditions are replaced with reflective ones. 
Our focus is on the extreme mass-ratio inspirals in which the central heavier 
black hole is replaced with an exotic compact object. 
We show, in this case, a modulation of the energy loss rate 
depending on the evolving orbital frequency occurs and leads to two different types of modifications to the 
gravitational wave phase evolution; the oscillating part directly corresponding to the 
modulation in the energy flux, and the non-oscillating part coming from 
the quadratic order in the modulation. This modification can be sufficiently large to detect with
future space-borne detectors.
\end{abstract}

\maketitle

\section{Introduction}
The nature of black hole (BH) event horizon is a focus of new frontiers of physics. 
The stringy paradigm of BHs predicts a little exotic nature of 
BH event horizon, which may appear as a modified boundary condition 
for perturbations around a BH spacetime~\cite{Skenderis:2008qn,Almheiri:2012rt}. 

If BHs in coalescing binary systems 
were replaced with slightly fat exotic objects like boson stars or gravastars~\cite{Liebling:2012fv,Mazur:2001fv,Visser:2003ge}, the effect may appear in the binary inspiral waveform through the tidal deformability or the spin-induced quadrupole moment~\cite{Wade:2013hoa,Cardoso:2017cfl,Sennett:2017etc,Krishnendu:2017shb,Kastha:2018bcr}. 
In this respect, 
constraints from the application to the LIGO-Virgo data have been already obtained~\cite{Johnson-McDaniel:2018uvs,Krishnendu:2019tjp,Narikawa:2021pak}. 
In future, space gravitational wave antenna missions such as LISA (Laser Interferometer Space Antenna) ~\cite{Audley:2017drz} will open a new window to search for exotic compact objects (ECOs). 
There are various proposals aiming at discriminating ECOs from ordinary BHs~\cite{Maselli:2017cmm,Cardoso:2019rvt,Guo:2019sns,Toubiana:2020lzd}. 

If the modification from the classical prediction of general relativity (GR) appeared 
only in the vicinity of the place where the event horizon existed in the BH case
(\textit{e.g.} area quantization Refs~\cite{Bekenstein:1974jk,Mukhanov:1986me}), 
the signal of modification would appear in a different way. One possible signature 
is the presence of echoes caused by the the waves reflected by the modified boundary 
condition~\cite{Cardoso:2016rao, Cardoso:2016oxy} (Also see \cite{Cardoso:2019apo, Agullo:2020hxe}. 
Some analyses of echoes in gravitational waves (GWs) just after binary coalescence suggest 
the presence of reflecting boundary for low frequency GWs near the horizon
~\cite{Abedi:2016hgu, Abedi:2017isz, Abedi:2020sgg}.
There have been many follow-up analyses about echo signal 
~\cite{Ashton:2016xff, Westerweck:2017hus, Nielsen:2018lkf, Lo:2018sep, Uchikata:2019frs, Abbott:2020jks} 
and most of them are 
negative to the presence of signal. In particular, the echo waveform 
should have an imprint of the angular momentum barrier around the BH if the modification 
is restricted to the boundary conditions near the horizon
~\cite{Nakano:2017fvh, Mark:2017dnq, Conklin:2019fcs, Maggio:2019zyv, Micchi:2019yze}. 
In our previous analysis ~\cite{Uchikata:2019frs}, we did not find any significant signal 
when we adopt such a physically motivated waveform as the template.  
However, it is also true that there seems some 
statistically meaningful signature which cannot be interpreted as random noises, 
if we follow the analysis method adopted in the original claim. 

The process of binary coalescence is highly dynamical, and therefore the modification from 
the prediction of classical GR might be a little more complicated than 
described by the modification to the boundary condition near the horizon. Here in this paper 
we argue that the best place to test the hypothesis of  near-horizon reflective boundary conditions should be extreme mass-ratio inspirals (EMRIs), 
which are one of the targets of LISA 
and other future space GW detectors
~\cite{Guo:2018npi, Luo:2015ght, Kawamura:2020pcg, Graham:2017lmg, Kuns:2019upi}. 

During the observation period, we can observe many cycles of GW oscillations. 
Thus, we can determine the phase evolution of EMRIs very precisely. 
The leading order of the phase evolution is determined by the secular variation of the orbital
parameters, under the adiabatic approximation. 
The modification to the boundary condition near the horizon is expected to modify the 
GWs emitted by the EMRI system and therefore the phase evolution.
Previously, there are several works to see the distinction of ECOs from BHs in the EMRI waveform in the contexts of tidal heating~\cite{Datta:2020rvo,Datta:2019epe,Datta:2019euh,Maselli:2017cmm}, tidal deformability~\cite{Pani:2019cyc} and area quantization~\cite{Agullo:2020hxe,Datta:2021row}.
We show that the effect appears in a rather distinctive way in the case of reflective boundary conditions, {\em i.e.}, the energy loss rate from the EMRI system suffers from a modulation caused by the interference among the waves directly emitted to the infinity and the reflected ones (There are some studies for the Schwarzschild case
\cite{Agullo:2020hxe,Cardoso:2019nis}. 
Furthermore,  
we heard a work by E. Maggio, M. van de Meent and P. Pani~\cite{Maggio:2021uge}, which independently obtained 
very similar results, after we almost finished writing this paper).
The energy loss rate is increased or decreased depending on the orbital frequency. 
The magnitude of this effect on the binary phase evolution is expected to be sufficiently large to detect. While the 
energy loss rate averaged over a sizable range of frequency turns out to be identical to the one in GR.  
Nevertheless, the cumulative effect on the EMRI phase evolution that appears non-linearly is also detectable. 

This paper is organized as follows. In Sec.\ref{sec:fundamentals}, we derive the rate of the energy loss
of a EMRI system with a hypothetical, reflective boundary, based on the black hole perturbation theory.
In Sec.\ref{sec:results}, we show some results of numerical calculation to estimate the impact of the
reflective boundary conditions to the evolution and the gravitational waves of EMRIs.
Finally, we discuss the correction to the number of cycles of gravitational waves from EMRIs, especially
the oscillating part which received little attention so far. We also give a brief discussion on the
detectability of the correction by future observation of LISA.
Throughout this paper we adopt the geometrized units with $G=c=1$. 

%%%%%%%%%%%%%%%%%%%%%%%%%%%%%%%%%%%%%%%%%%%%%%%%%%%%%%%%%%%%%%%%%
\section{Energy loss rate with modified boundary} \label{sec:fundamentals}
%%%%%%%%%%%%%%%%%%%%%%%%%%%%%%%%%%%%%%%%%%%%%%%%%%%%%%%%%%%%%%%%%
In GR Teukolsky radial function that satisfies 
\begin{equation}
    \Delta^{-s} \frac{d}{dr} \left(\Delta^{s+1} \frac{dR}{dr}\right)+V(r) R =0\,,
\label{eq:Teukolsky}
\end{equation}
with an appropriate potential $V(r)$~\cite{Teukolsky:1974yv}, 
takes the following asymptotic forms of solutions,
\begin{align}
&R=\left\{
\begin{array}{ll}
r^{-1-s\pm s}e^{\mp i \omega r^*}\,,\quad & \mbox{for}~ r^*\to +\infty\,,\cr
\Delta^{(-s\pm s)/2}e^{\pm i k r^*}\,,\quad & \mbox{for}~ r^*\to -\infty\,,
\end{array}
\right.
\end{align}
where $k=\omega -m\Omega_H$, $r^*=\int dr (r^2+a^2)/\Delta$, $\Omega_H=a m/2Mr_+$, $M$ and $a$ are the mass and the Kerr parameter of the BH, and $r_+$ is the horizon radius in the Boyer-Lindquist coordinates. 
We can define the reflection and transmission coefficients ${\cal R}$ and ${\cal T}$ 
for the incident in-going wave from 
infinity as 
\begin{align}
&R_\textrm{in}\!=\!\left\{\!\!
\begin{array}{ll}
r^{-1}e^{-i \omega r^*}
+{\cal R} r^{-1-2s}e^{i \omega r^*}\!, & \mbox{for}~ r^*\to +\infty\,,\cr
{\cal T} \Delta^{-s}e^{- i k r^*}\!, & \mbox{for}~ r^*\to -\infty\,.
\end{array}
\right. \label{eq:in-going-R}
\end{align}
Similarly, for the wave coming from the past horizon, we have 
\begin{align}
&R_\textrm{up}=\left\{
\begin{array}{ll}
\tilde{\cal T}r^{-1-2s}e^{i \omega r^*}\,,\quad & \mbox{for}~ r^*\to +\infty\,,\cr
e^{i k r^*}+\tilde {\cal R} \Delta^{-s}e^{- i k r^*}\,,\quad & \mbox{for}~ r^*\to -\infty\,.
\end{array}
\right. \label{eq:up-coming-R}
\end{align}
The Wronskian relation tells that 
\begin{align}
  W(R_1,R_2):= \Delta^{s+1} \left(R_1 \frac{dR_2}{dr} -R_2 \frac{dR_1}{dr}\right) 
\end{align}
is constant, if both $R_1$ and $R_2$ are solutions of Eq.~\eqref{eq:Teukolsky}, which implies the relation 
\begin{align}
    \tilde {\cal T}=\frac{2Mr_+ k -is(r_+-M)}{\omega}{\cal T}\,.
\end{align}

We consider a modified boundary condition such 
that implies 
\begin{align}
    \frac{\alpha^+}{\alpha^-}=R_{\rm b}\frac{\epsilon_-}{\epsilon_+}\,, 
\end{align}
where $\alpha^\pm$ are the coefficients in the asymptotic form of the 
radial function at $r^*\to -\infty$, 
\begin{align}
    R\to \alpha^+ e^{ikr^*} +\alpha^- \Delta^{-s} e^{-ikr^*}\,, 
    \label{eq:modifiedAsymptoticForm}
\end{align}
and $\epsilon_\mp$ are the coefficients that translate the bare amplitude of the radial function $R(r)$ 
into the energy fluxes into/out of the horizon. 
The explicit expressions for $\epsilon_\mp$ are given by~\cite{Teukolsky:1974yv}
\begin{eqnarray}
\epsilon_{-}^2 &=&
\frac{256(2Mr_+)^5 (k^2+4\epsilon^2) (k^2+16\epsilon^2) k \omega^3}
{|C_{SC}|^2}\,, \\
\epsilon_{+}^2 &=&
\frac{\omega^3}{k(2Mr_+)^3(k^2+4\epsilon^2)}, 
\end{eqnarray}
with 
\begin{eqnarray}
\epsilon &=& \frac{\sqrt{M^2-a^2}}{4Mr_+}, \\
|C_{SC}|^2 &=&
\left[ (\lambda+2)^2 + 4a\omega m - 4a^2\omega^2 \right]\cr
&&\times \left[ \lambda^2 + 36a\omega m - 36a^2\omega^2 \right]
+ 144 \omega^2 ( M^2  - a^2 )\nonumber \\ &&
+ 48 a\omega (2\lambda + 3) ( 2a\omega - m )
\,,
\end{eqnarray}
where $\lambda$ is the separation constant between the radial and angular equations.  
$R_{\rm b}$ is a complex number whose squared absolute value $|R_{\rm b}|^2$ gives the 
reflection rate at the boundary. 
$\epsilon_\mp^2$ is negative when $\omega$ and $k$ have the opposite signature, which is the condition for the superradiance, {\em i.e.}, $0<\omega<m\Omega_H$ for a positive $m$. 
We denote the Green's function for the original GR problem as 
\begin{align}
    G(r,r')=\frac{1}{W(R_\textrm{in},R_\textrm{up})}
     & \Bigl\{ R_\textrm{in}(r) R_\textrm{up}(r')\theta(r'-r)\cr
     & \!\!\!\! +R_\textrm{up}(r) R_\textrm{in}(r')\theta(r-r')\Bigr\}\,.
\end{align}
The Green's function for the modified boundary condition would be 
obtained by simply replacing $R_\textrm{in}$ with 
\begin{align}
    \tilde R_\textrm{in}:= R_\textrm{in} +\beta R_\textrm{up}
                  = \beta e^{ikr^*}\!\!\! +({\cal T}+\beta \tilde{\cal R}) \Delta^{-s} e^{-ikr^*}\!\!,
\end{align}
where $\beta$ is determined as such that $\tilde R_{in}$ is proportional to the 
asymptotic form given in Eq.~\eqref{eq:modifiedAsymptoticForm}. More explicitly,  
$\beta$ is given by 
\begin{align}
\beta={\cal T} R_{\rm b} \frac{\epsilon_-}{\epsilon_+}
   \left(1-\tilde {\cal R}R_{\rm b}\frac{\epsilon_-}{\epsilon_+}\right)^{-1}\,,
\end{align}
in terms of $R_{\rm b}$. 

Under the ordinary outgoing boundary conditions in GR, we would have 
\begin{align}
&R=\left\{
\begin{array}{ll}
Z_\infty \tilde{\cal T} r^{-1-2s}e^{i \omega r^*}\,,\quad & \mbox{for}~ r^*\to +\infty\,,\cr
Z_H {\cal T} \Delta^{-s}e^{- i k r^*}\,,\quad & \mbox{for}~ r^*\to -\infty\,, 
\end{array}
\right.
\end{align}
and the coefficients $Z_\infty$ and $Z_H$ are given by 
\begin{align}
    Z_{\infty/H} & = \frac{1}{W(R_\textrm{in},R_\textrm{up})}\int S(r) R_\textrm{in/up}(r) dr\,,
    \label{eq:asymptotic_amplitude}
\end{align}
where $S(r)$ is the appropriate source function \cite{Sasaki:2003xr}. 
The energy fluxes to the infinity and to the horizon are, respectively, given by 
\begin{align}
    F^{(\infty)}=\frac{|\tilde{\cal T}|^2|Z_{\infty}|^2}{4\pi\omega^2}\,,\qquad 
    F^{(H)}=\frac{\epsilon_-^2 |{\cal T}|^2 |Z_{H}|^2}{4\pi\omega^2}\,.
    \label{eq:flux}
\end{align}

\begin{figure*}[thbp]
\begin{center}
\includegraphics[width=0.4\linewidth]{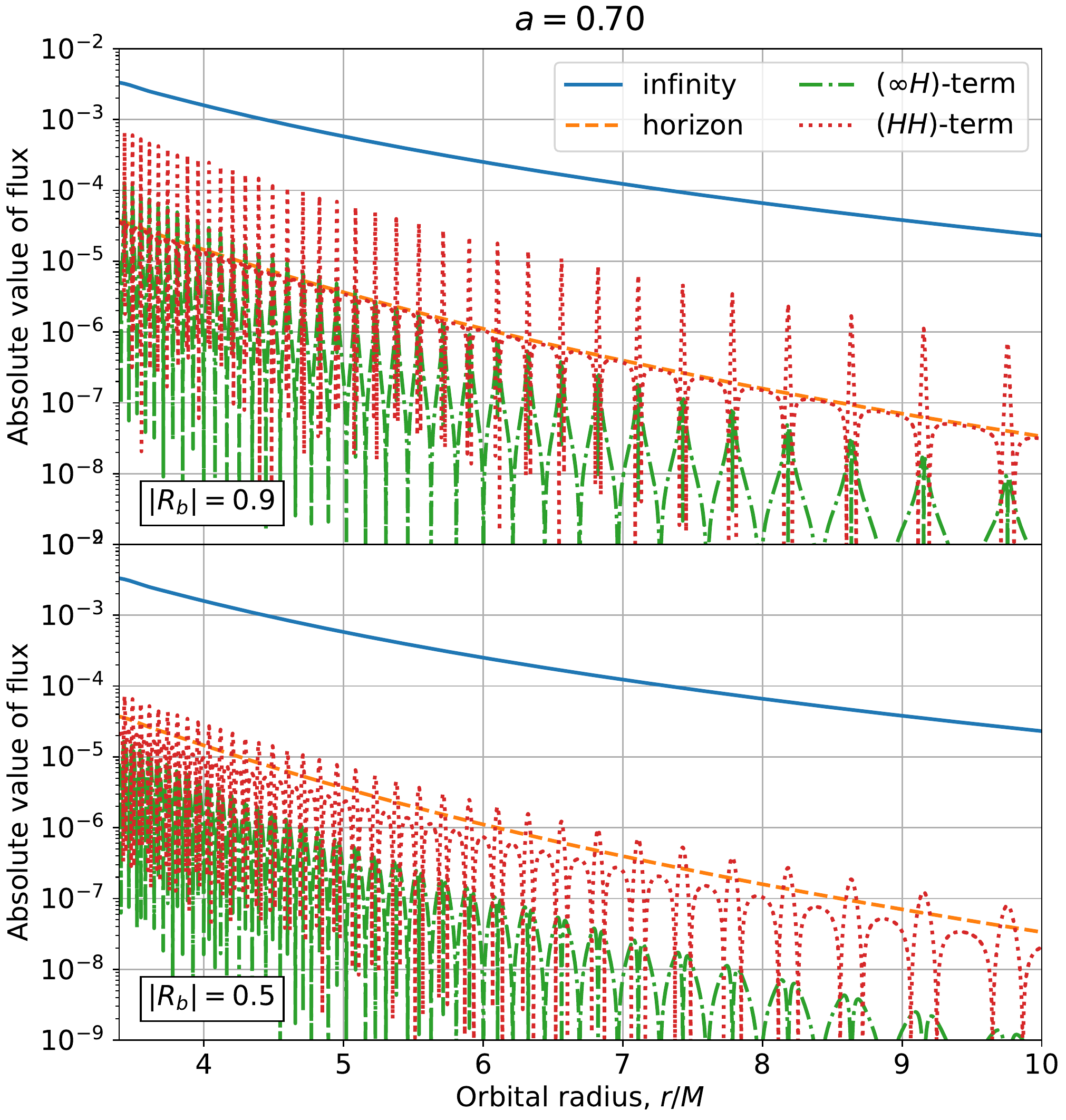} \hspace{5mm}%\\%
\includegraphics[width=0.4\linewidth]{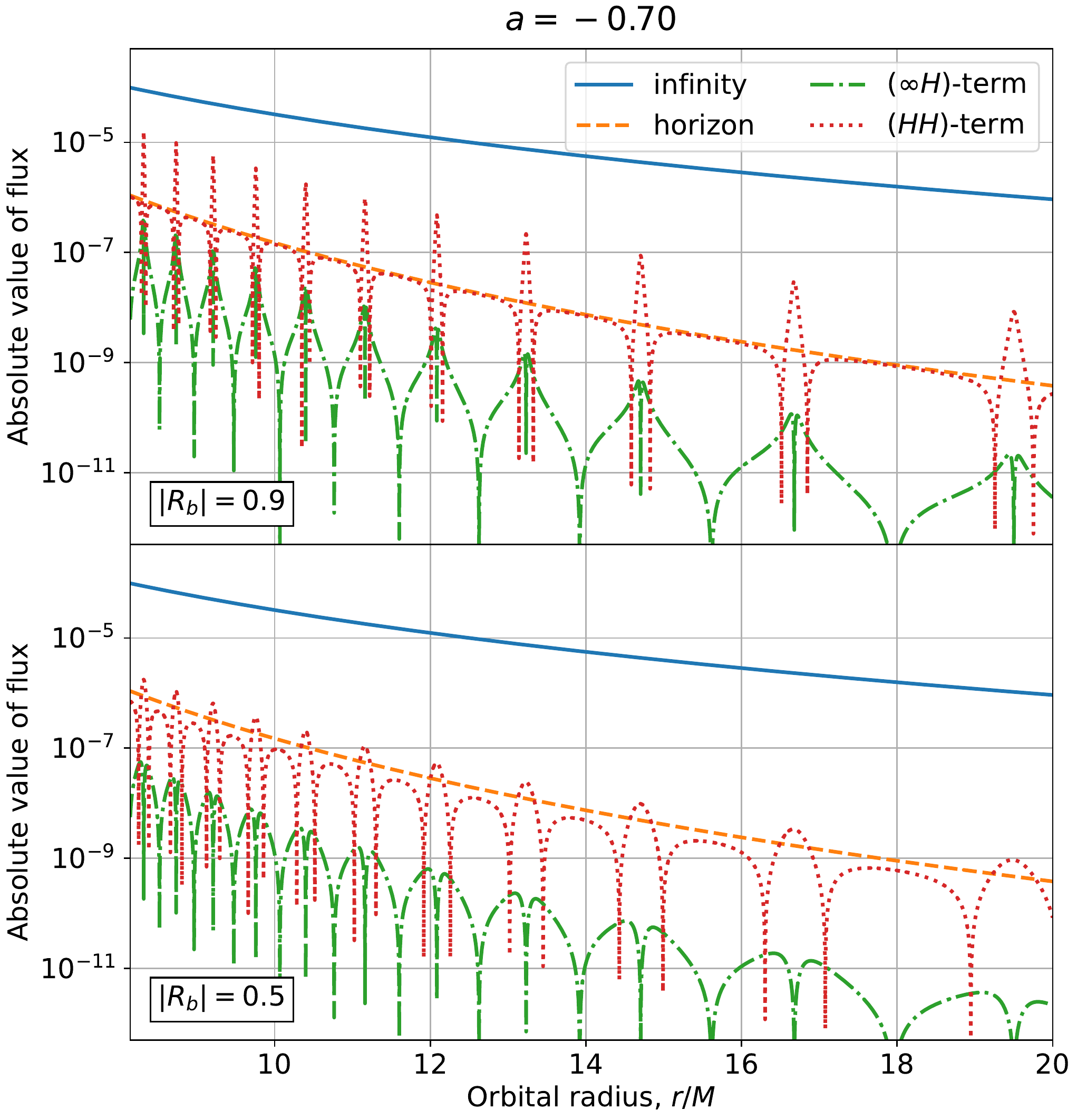}
\end{center}
\vspace{-5mm}
\caption{Energy flux and correction to energy flux due to the modification 
of the inner boundary condition. Here we consider the $(l,m)=(2,2)$ mode for 
circular equatorial orbits, and set $|a|=0.7M$, $r^*_{\rm b}=-500M$,
$R_{\rm b}=|R_{\rm b}|e^{-2ikr^*_{\rm b}}$. We show the results for four cases,
$(a,|R_{\rm b}|)=(0.7M,0.9)$ (top, left),  $(0.7M,0.5)$ (bottom, left), 
$(-0.7M,0.9)$ (top, right), and $(-0.7M,0.5)$ (bottom, right).}
\label{fig:flux}
\end{figure*}

For the modified boundary conditions, we have
\begin{align}
&    F^{(\infty)}_{\rm mod}=\frac{|\tilde {\cal T}|^2|(Z_{\infty}+\beta Z_H)|^2}{4\pi\omega^2}\,,\cr
&    F^{(H)}_{\rm mod}=
     \frac{\left(\epsilon_-^2 |{\cal T}+\beta\tilde {\cal R}|^2|
        -\epsilon_+^2|\beta|^2 \right)|Z_{H}|^2}{4\pi\omega^2}\,.
    \label{eq:modified_flux}
\end{align}
Then, the difference of the total energy flux from the GR case is computed as 
\begin{align}
  \delta F^\textrm{(tot)}&:= 
   \left(F^{(\infty)}_{\rm mod}+F^{(H)}_{\rm mod}\right)-
    \left(F^{(\infty)}+F^{(H)}\right)\cr
    & =\delta F^{\infty H} +\delta F^{HH}\,,
\label{eq:difF}
\end{align}
where
\begin{align}
\delta F^{\infty H} = &
 \frac{|4Mr_+ k-2is(r_+-M)|}{\omega}\cr
 &\quad\times \sqrt{|F^{(\infty)} F^{(H)}|} \Re\left(
 \frac{Z_\infty^{*} Z_H \beta}{|Z_\infty Z_H \epsilon_-|}\right),
\cr
\delta F^{HH} = &
 2F^{(H)} \Re\left(
         \frac{\tilde{\cal R}\beta}{{\cal T}}\right)\,.
         \label{eq:modifiedFlux}
\end{align} 
Here, we used the relation 
\begin{align}
    \epsilon_+^2-\epsilon_-^2|\tilde{\cal R}|^2=|\tilde{\cal T}|^2\,,
\end{align}
which follows from the energy flux conservation. Respective terms in $\delta F^\textrm{(tot)}$ 
oscillate depending on the phases in $\Re$, and their amplitudes of oscillation 
$\overline{\delta F}{}^{\infty H}$ and $\overline{\delta F}{}^{H H}$ 
are simplified as 
\begin{align}
  \frac{\overline{\delta F}{}^{\infty H}}{\sqrt{|F^{(\infty)} F^{(H)}|}}
  = &
  \frac{|2Mr_+ k -is(r_+-M)|^2|\beta|^2}{\omega^2\epsilon_-^2}\cr
     \qquad\qquad\qquad = & \frac{1-|\hat{\cal R}|^2}
       {|1-\hat{\cal R} R_{\rm b} |^2} |R_{\rm b}|^2 \,,\cr
   \frac{\overline{\delta F}{}^{HH}}{|F^{(H)}|}
  = &
  \frac{|\beta\tilde{\cal R}|^2}{|{\cal T}|^2}
     = \frac{|\hat{\cal R} R_{\rm b}|^2}{|1-\hat{\cal R} R_{\rm b} |^2}\,, 
\label{eq:modA}
\end{align}
where we have defined
\begin{align}
    \hat{\cal R}:= \frac{\epsilon_-}{\epsilon_+}\tilde{\cal R}\,. 
\end{align}
$|\hat{\cal R}|^2$ represents the physical reflection rate by the angular momentum potential barrier. 
At first glance, in Eq.~\eqref{eq:difF}, $\delta F^{\infty H}$ seems dominant since $F^{(\infty)}\gg F^{(H)}$, {\em i.e.}, $F^{(H)}$ is 2.5 higher order in the post-Newtonian expansion. (Strictly speaking, logarithmic correction is also associated.) 
However, for modes with small $M\omega$, typical for the inspiral phase, $|\hat{\cal R}|$ should be 
close to unity. Thus, $\overline{\delta F}{}^{\infty H}$ is suppressed. 
In fact, since $|\hat{\cal R}|$ gets closer to unity for a smaller value of $M\omega$, the amplitude of $\delta F^{\infty H}$ decreases more rapidly than $\delta F^{HH}$ toward smaller values of $M\omega$, as one can see in Fig.~\ref{fig:flux}. 
In contrast, the denominators in Eqs.~\eqref{eq:modA} can get close to 
zero for $|R_{\rm b}|\approx 1$ depending on the phase. The phase of $\hat{\cal R} R_{\rm b}$ will vary depending 
on the frequency and typically the period of one cycle in $\omega$, $\Delta\omega$, would be approximately a constant roughly 
estimated by $2\Delta\omega |r^*_{\rm b}|\approx 2\pi$, where $r^*_{\rm b}$ is the hypothetical boundary in tortoise coordinate. 
As long as we assume the boundary is placed at around the Planck distance as in the models of echo signal, we obtain $|r^*_{\rm b}/M|=O(500)$ robustly. 
Hence, it is expected that the phase will vary by much more than $2\pi$ during the inspiral.

An interesting observation is that the rapid frequency dependence in the expressions~\eqref{eq:modifiedFlux} comes only from the phase of $\beta$. 
If we neglect the other frequency dependence, 
$\delta F^{\infty H}$ and $\delta F^{HH}$ averaged over a certain frequency range take the form of 
\begin{equation}
    \Re\int d\omega \frac{c_1 e^{i r_{\rm b}^* \omega}}{1-c_2 e^{i r_{\rm b}^* \omega}}\,,
\label{eq:approximation}
\end{equation}
with some constants $c_1$ and $c_2$. This form of integral over one period of the cycle $\Delta\omega$ exactly vanishes. 

Here, we should worry about the presence of super-radiant instability in the ergo-region~\cite{Friedman:1978,Vilenkin:1978uc,Brito:2015oca}. 
If $|R_{\rm b}|$ is close to unity, 
the Kerr BH mimicker will become unstable. When we increase $|R_{\rm b}|$, 
a super-radiant instability mode appears when  
\begin{align}
    R_{\rm b}=\hat{\cal R}^{-1}\,,
\label{eq:AppearanceSR}
\end{align}
is satisfied for some real value of $\omega$. We should recall that 
$|\hat{\cal R}|>1$ for modes with $\omega < m\Omega_H$. 
Therefore, the condition \eqref{eq:AppearanceSR} can hold even when $|R_{\rm b}|<1$. 
The difference in 
the total energy flux $\delta F$ diverges there, as we can see from the vanishing of the 
denominators in Eqs.~\eqref{eq:modA}. Therefore, the analysis presented here is valid only when $R_{\rm b}$ is arranged to 
evade the ergo-region instability, as a natural requirement.  

Interestingly, even if we consider the case of perfect reflection with $|R_{\rm b}|=1$, 
the instability can be evaded if the phase of $R_{\rm b}$ is appropriately arranged 
not to satisfy the condition \eqref{eq:AppearanceSR} for any amplitude $|R_{\rm b}|\leq 1$ and 
for any value of $\omega$. However, as long as we consider some physically motivated boundary 
at the Planck distance from the horizon, $R_{\rm b}$ should contain an overall factor like 
$e^{2i\omega r^*_{\rm b}}$, corresponding to the round-trip distance to the boundary. 
Therefore, absence of super-radiant modes will not be 
compatible with the perfect reflection. 

%%%%%%%%%%%%%%%%%%%%%%%%%%%%%%%%%%%%%%%%%%%%%%%%%%%%%%%%%%%%%
\section{Results} \label{sec:results}
%%%%%%%%%%%%%%%%%%%%%%%%%%%%%%%%%%%%%%%%%%%%%%%%%%%%%%%%%%%%%
In this following, to evaluate the impact of the modified boundary conditions to the evolution of an EMRI,
we compute the energy fluxes of the gravitational radiation from the EMRI system.
To evaluate both the unmodified and modified fluxes in Eqs.~(\ref{eq:flux}) and (\ref{eq:modified_flux}), 
in this work, we calculate the two homogeneous solutions satisfying Eqs.~(\ref{eq:in-going-R}) and
(\ref{eq:up-coming-R}) by using the Sasaki-Nakamura formalism~\cite{Sasaki:1981kj}. We also need to calculate
the asymptotic amplitudes, Eq.~\eqref{eq:asymptotic_amplitude}, including the source term of the
inhomogeneous Teukolsky equation. The explicit formulas of the amplitudes for general orbits are given 
in the literature, {\em e.g.}, Ref.~\cite{Sasaki:2003xr}.
Since the modified boundary condition depends on the unknown physics, we set the reflection rate on
the boundary as $R_{\rm b}=|R_{\rm b}|e^{-2ikr^*_{\rm b}}$ with
$|R_{\rm b}|=(0.50, 0.90)$, by hand.
The phase factor simply reflects the distance to the boundary, and no additional phase shift depending on the frequency is assumed. 

EMRIs are likely to be eccentric at birth. However, since the eccentricity rather quickly decays, some fraction of EMRIs will become nearly circular before the plunge. 
Furthermore, EMRIs in circular orbits would be more suitable for capturing the effect of the reflective boundary, because of the simplicity of the waveform. As for the orbital inclination, since the horizon in-going flux is suppressed for a large inclination ({\it e.g.}, see \cite{Sago:2015rpa}), EMRIs in equatorial orbits are more suitable for the present purpose. Therefore, we select equatorial circular EMRIs as a representative case.

\begin{figure}[htbp]
\includegraphics[width=0.8\linewidth]{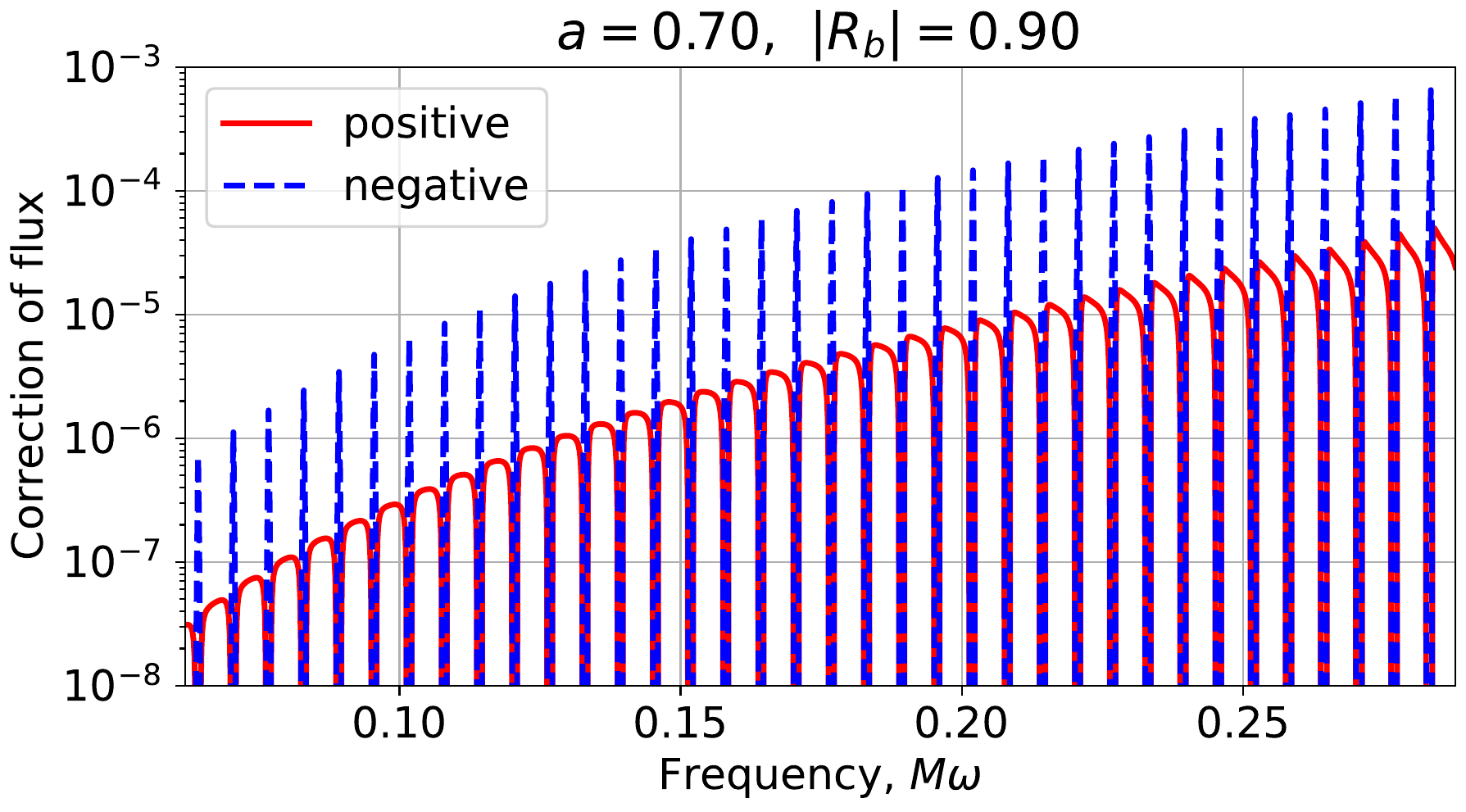} \\%
\includegraphics[width=0.8\linewidth]{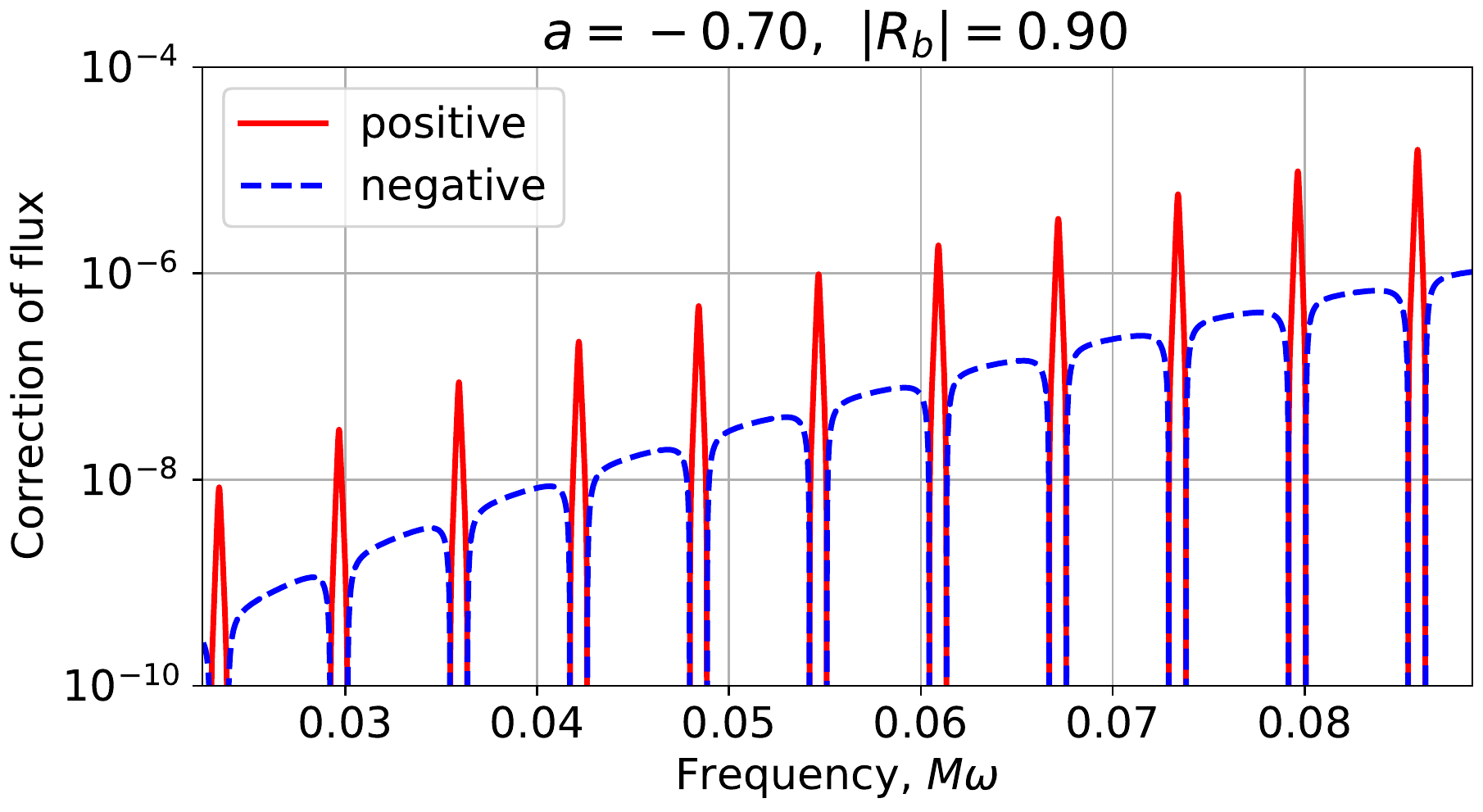}%
\caption{Total correction to energy flux due to the modification
of the inner boundary condition, for equatorial and circular orbits with $a=0.7M$ (upper panel) and $a=-0.7M$ (lower). We fix $(l,m)=(2,2)$, $r^*_{\rm b}=-500M$,
$R_{\rm b}=0.9e^{-2ikr^*_{\rm b}}$.
The horizontal axis is the GW frequency $\omega$, and vertical axis is the flux divided by the mass ratio $\mu/M$.}
\label{fig:corr_flux}
\end{figure}

In Fig.\ref{fig:flux}, we show 
the fluxes of $(l,m)=(2,2)$ mode for the two different values of $|R_{\rm b}|$
with $a=\pm 0.7M$, $r^*_\textrm{b}=-500M$. (Plus sign of $a$ means that the orbit is co-rotating.)
Here, we note that, since the GW frequency $\omega$ 
is less than the BH horizon angular frequency $\Omega_H$, the horizon flux is negative (super-radiant).
In the same plots, we also show the correction terms to the modified flux. 
We can see that, as mentioned before,
$\delta F^{\infty H}$ 
(green dash-dot lines in the plots) is suppressed compared to $\delta F^{HH}$.
We can also find several peaks for all cases, which correspond to
the resonance induced by the confinement of GWs between the angular 
potential barrier and the reflective surface. These peaks become higher and narrower
when $|R_{\rm b}|$ gets close to unity.
The interval between two successive peaks is roughly estimated by 
$\Delta\omega\sim\pi/|r_\textrm{b}^*|\sim 0.0063/M$ (see also Fig.\ref{fig:corr_flux}). 
The interval depends on the spin parameter of the central BH 
through the location of the angular momentum potential barrier, 
but its change due to spin is tiny of $O(M)$, which is much smaller 
than $|r^*_\textrm{b}|=500M$.

The plots in Fig.\ref{fig:corr_flux} show the total modification to the flux, 
$\delta F$, for $a=\pm 0.7M$ and $|R_{\rm b}|=0.90$ cases. %in Fig.\ref{fig:flux}. 
To specify the direction of the modification in the semi-logarithmic scale, we plot 
the positive contribution with red solid curves, and the negative one with blue dashed curves. A positive modification of the flux makes the evolution of the inspiral faster,
while a negative one does it slower.
We can confirm that the peaks are placed with an equal separation in $\omega$, and 
the signature of peaks is opposite between the co-rotating and counter-rotating cases.

\begin{figure*}[htbp]
\includegraphics[width=0.40\linewidth]{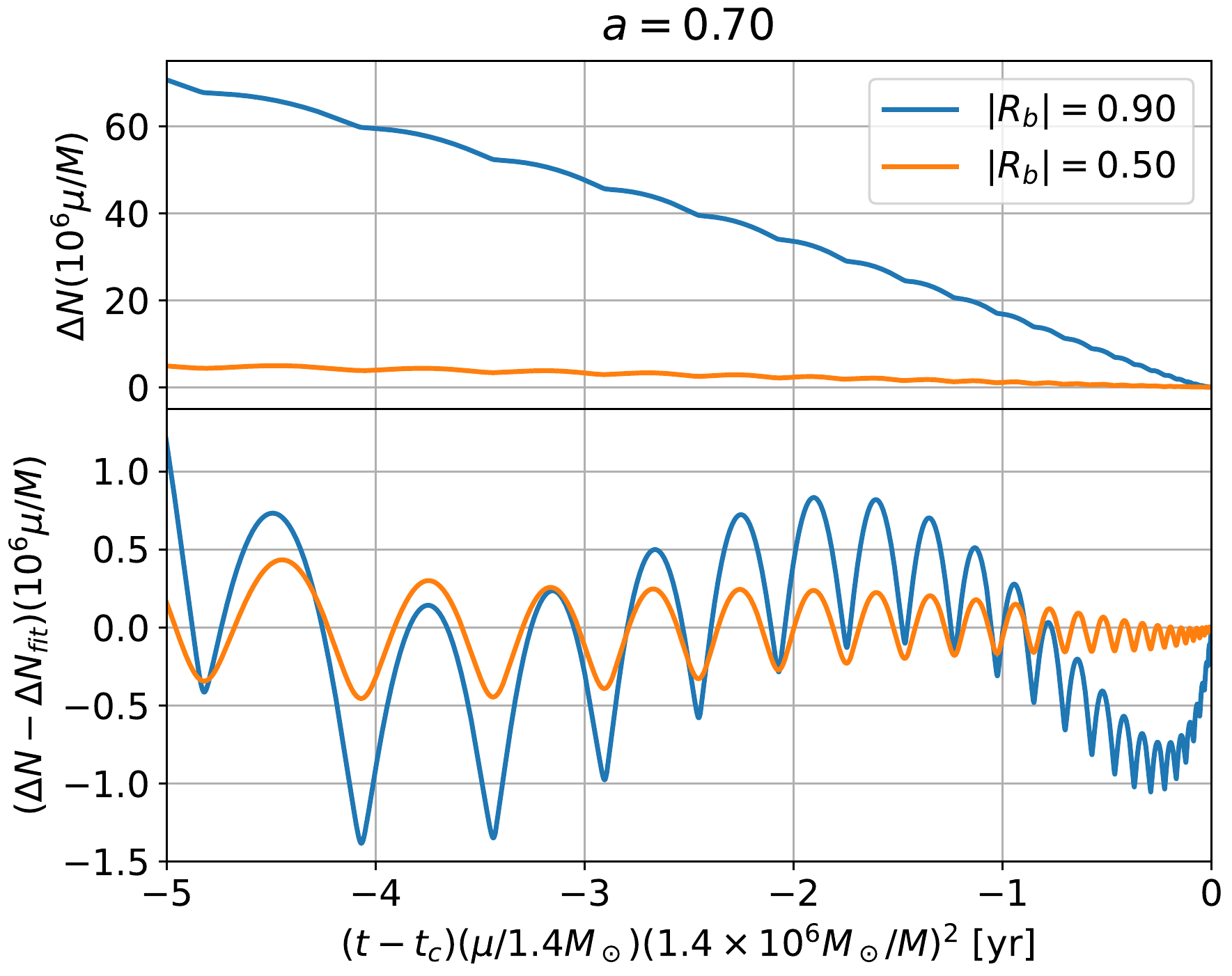}\hspace{5mm}%
\includegraphics[width=0.40\linewidth]{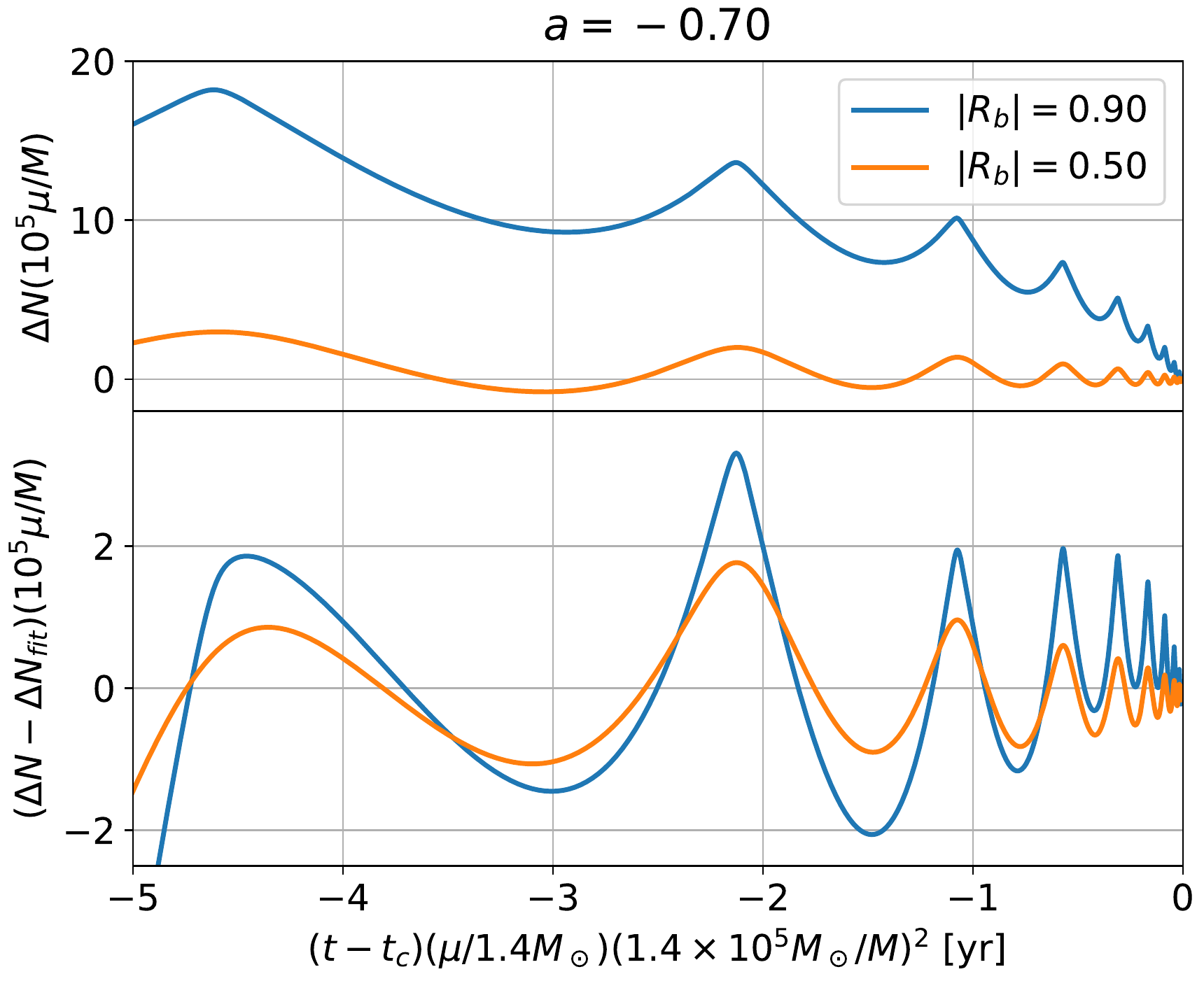}%
\caption{
Time dependence of the cycle of GW emitted by a quasi-circular, equatorial EMRI
with the component masses, $(\mu,M)=(1.4, 1.4\times 10^6)M_\odot$ for $a=0.7$ (left) and $(\mu, M)=(1.4, 1.4\times 10^5)M_\odot$ for $a=-0.7$ (right).
Here we consider two boundary conditions, $R_{\rm b}=0.9e^{-2ikr^*_{\rm b}}$ and 
$0.5e^{-2ikr^*_{\rm b}}$ with $r^*_{\rm b}=-500M$. 
The upper panels show the corrections to the number of cycles during five years for the orbiting object to reach the ISCO radius, induced by the modification of
the inner boundary condition.
The lower panels show the oscillating parts of the corrections extracted from $\Delta N$ by subtracting the smoothly changing part.
During the last five years before reaching the ISCO, the GW
frequency sweeps from 3.04 mHz
to 6.64 mHz for the left panel, and 
from 5.27 mHz
to 20.5 mHz for the right panel.}
\label{fig:Ncycle_vs_t2}
\end{figure*}

To evaluate the influence of the modification on the GW waveform clearly, we 
consider the number of cycles of GWs from a quasi-circular, equatorial EMRI, 
defined by
\begin{equation}
N = \int_{r(t)}^{r_{\rm ISCO}} \frac{\Omega_\phi}{\pi} \frac{dE/dr}{dE/dt} dr,
\end{equation}
where $\Omega_\phi$, $E$ and $r_{\rm ISCO}$ are the orbital angular velocity, the 
specific energy of the particle and the radius of the inner-most stable circular orbit (ISCO), from which the cycle 
is accumulated. 
From the balance argument of the energy, the energy loss of the
particle, $dE/dt$ can be estimated by the flux of the GW emitted
by the EMRI system, $F$, as $dE/dt=-F$. For the purpose of estimating the order of magnitude of the effect, here we take into account only the leading order contribution of $F$ due to $(l,m)=(2,\pm 2)$ mode. (Notice that the contribution from $m=-2$ mode is the same as $m=2$ because of the symmetry under $(\omega,m)\to(-\omega,-m)$.)
We denote the number of cycles with the unmodified and modified fluxes as $N(t)$ and $N_\textrm{mod}(t)$, respectively.
In the upper panels in Fig.\ref{fig:Ncycle_vs_t2}, we show the correction to the number of cycles, $\Delta N:=N_\textrm{mod}-N$, for five years before reaching the ISCO. 
Here, we consider two cases of EMRIs: one with the component masses, $(\mu, M)=(1.4,1.4\times 10^6)M_\odot$ and $a=0.7$ (left), and the other with $(1.4, 1.4\times 10^5)M_\odot$ and $a=-0.7$ (right).
The correction due to the boundary with $R_{\rm b}=0.9e^{-2ikr^*_\textrm{b}}$ ($0.5e^{-2ikr^*_\textrm{b}}$) 
is accumulated up to $O(10)$ ($O(1)$) for both cases, which will affect the observation of EMRIs by LISA.
We briefly discuss it in the following section.

%%%%%%%%%%%%%%%%%%%%%%%%%%%%%%%%%%%%%%%%%%%%%%%%%%%%%%%%%%%%%
\section{Discussion} \label{sec:discussion}
%%%%%%%%%%%%%%%%%%%%%%%%%%%%%%%%%%%%%%%%%%%%%%%%%%%%%%%%%%%%%
The correction of the number of cycles shown in the previous section,
$\Delta N$, is divided into two parts: one is the oscillating part and the other is the smoothly changing part. 
In the lower panel in Fig.\ref{fig:Ncycle_vs_t2}, we show the plots to emphasize the oscillating part of $\Delta N$ by subtracting its smoothly changing part. The smoothly changing part is obtained by fitting $\Delta N$ in the plotted interval (five years) by a third order polynomial function in the form of $a_1 (t-t_c) + a_2(t-t_c)^2 + a_3(t-t_c)^3$, where $t_c$ is the time to pass the ISCO radius.
In many previous studies, basically something like the latter part was discussed. 
However, in the present model the energy flux averaged over frequency does not deviate from the case of GR. In fact, $\Delta N$ at a large $|t-t_c|$ (outside of these plots) can be explained by the shift of $t_c$ and constant phase.
However, as shown in Fig.~\ref{fig:Ncycle_vs_t2}, definitely there exists non-oscillating part of the phase shift. This shift arises because the average is not taken over the frequency but over the time. The frequency ranges corresponding to the peaks in $\delta F$ is swept slowly (quickly) compared with the other parts for the co-rotating (counter-rotating) case. This effect appears 
at the second order of $\delta F$, and $\Delta N$ due to this effect 
is roughly estimated by $\int f \left<(\delta F/F)^2\right> dt$. 
The dependence of the frequency average of $(\delta F/F)^2$, $\left<(\delta F/F)^2\right>$, on $R_{\rm b}$ is  
proportional to $|\hat {\cal R} R_{\rm b}|^2/(1-|\hat {\cal R} R_{\rm b}|^2)$, which explains the ratio between the magnitudes of $\Delta N$ for 
the two cases with $|R_{\rm b}|=0.5$ and $0.9$. In the plotted range, the change of the frequency is so tiny that $\Delta N$ looks proportional to $t-t_c$, which would be consistent with the above estimate. 
In the post-Newtonian regime, the effect should behave as 5th post Newtonian correction, reflecting the scaling of $\left<(\delta F/F)^2\right>$. 
The magnitude of this correction to the phase 
is larger compared with the oscillating part, but it is not clear if we can say that this signal is really uniquely caused by the reflective boundary.  

For the oscillating part, we can estimate the amplitude of 
oscillation of $\Delta N$, $\overline{\Delta N}$, and 
the oscillation period $T$, using the same approximation adopted in Eq.~\eqref{eq:approximation}, as
\begin{align}
&\overline{\Delta N}\approx T \frac{F^{(H)}|\hat{\cal R} R_{\rm b}|}{8r_{\rm b}^* F}\,,
\quad T\approx \frac1{r_{\rm b}^*}\left(\frac{df}{dt}\right)^{-1}\,. 
\end{align}
Here, we assume $\delta F\approx \delta F^{HH}$. 
For a fixed observation period, $T$ should be at most comparable to the observation period, in order to detect the oscillations. Once $T$ is fixed, $\overline{\Delta N}$ is maximized by choosing the orbit close to the inner-most stable orbit and reducing the mass of the central BH. From these considerations, we chose 
the mass parameters in Fig.~\ref{fig:Ncycle_vs_t2}. 
These oscillations of the phase shift, which are periodic in terms of the GW frequency, should be very unique for the reflective near-horizon boundary, and it will be a smoking gun. 

Finally, to roughly prove the detectability of the phase shift, we estimate the 
maximum match between the modified GW waveform and the normalized unmodified GR template with the time and phase to pass the ISCO, $\mu$ and $M$ of the latter varied.  
The noise spectrum density of LISA is taken from Ref.~\cite{Robson:2018ifk}.
For $(\mu,M) = (1.4, 1.4\times 10^6) M_{\odot}$ and $R_\textrm{b}=0.01$, we obtain
${\cal M}=0.327$, which demonstrates the detectability of 
the reflective boundary even for a tiny reflection rate of $O(0.01)$.

\acknowledgments
We would like to thank anonymous referees for their useful comments to improve
the paper, especially the discussion on the detectability.
This work is supported by JSPS Grant-in-Aid for Scientific Research JP17H06358 (and also JP17H06357), as a part of the innovative research area, ``Gravitational wave physics and astronomy: Genesis'', and also by JP20K03928. 
NS also acknowledges support from JSPS KAKENHI Grant No. JP21H01082.
\bibliography{bib_ECObin}

\end{document}